# van der Waals Bonded Co/h-BN Contacts to Ultrathin Black Phosphorus Devices

Ahmet Avsar[†,‡,⊥], Jun Y. Tan[†,‡], Luo Xin[†,§], Khoong Hong Khoo[+], Yuting Yeo[†,‡], Kenji Watanabe[∥], Takashi Taniguchi[∥], Su Ying Quek[†,‡,*], Barbaros Özyilmaz[†,‡,*]

[†] Centre for Advanced 2D Materials, National University of Singapore, 117542, Singapore

[‡] Department of Physics, National University of Singapore, 117542, Singapore

[⊥] Electrical Engineering Institute and Institute of Materials Science and Engineering, École Polytechnique Fédérale de Lausanne (EPFL), Lausanne CH-1015, Switzerland

[§] Department of Applied Physics, The Hong Kong Polytechnic University, Hung Hom, Kowloon, Hong Kong, P. R. China

[+] Institute of High Performance Computing, 1 Fusionopolis Way, #16-16 Connexis, Singapore 138632, Singapore

[∥] National Institute for Materials Science, 1-1 Namiki, Tsukuba 305-0044, Japan

Corresponding Authors

*Email: barbaros@nus.edu.sg (experiment) ; phyqsy@nus.edu.sg (theory)

ABSTRACT: Due to the chemical inertness of 2D hexagonal-Boron Nitride (h-BN), few atomic-layer h-BN is often used to encapsulate air-sensitive 2D crystals such as Black Phosphorus (BP). However, the effects of h-BN on Schottky barrier height, doping and contact resistance are not well known. Here, we investigate these effects by fabricating h-BN encapsulated BP transistors with cobalt (Co) contacts. In sharp contrast to directly Co contacted p-type BP devices, we observe strong n-type conduction upon insertion of the h-BN at the Co/BP interface. First principles calculations show that this difference arises from the much larger interface dipole at the Co/h-BN interface compared to the Co/BP interface, which reduces the work function of the Co/h-BN contact. The Co/h-BN contacts exhibit low contact resistances (~ 4.5 k$\Omega$), and are Schottky barrier free. This allows us to probe high electron mobilities (4,200 cm$^2$/Vs) and observe insulator-metal transitions even under two-terminal measurement geometry.

Similar to graphite, black phosphorus is a single elemental material and its monolayers are held together by weak Van der Waals forces. It has thickness dependent direct band gap, varies from ~ 2 eV in monolayer to ~ 0.3 eV in bulk which cover the electromagnetic spectrum from visible to infrared portion[1]. While ultra-thin BP degrades under ambient conditions in a matter of minutes[2], the recent development of various encapsulation schemes make BP air-stable over months[3–5]. BP devices employing encapsulating h-BN layers exhibit the best electronic quality including the observation of directional dependent mobilities as high as 45,000 (5,200) cm$^2$/Vs at low (room) temperatures[6] and quantum oscillations[7].

Hexagonal Boron Nitride (h-BN) is a chemically inert layered material and it is free of pin holes and dangling bonds[8]. It plays different key roles in the optimization of 2D materials for future electronic applications. It has been used as a high quality substrate to enhance the electronic mobilities of 2D materials such as graphene[8], $MoS_2$[9], BP[7] and InSe[10]. Its ultra-thin layers have been also utilized as an encapsulating material to prevent the degradation of air-sensitive materials[5] and as a tunnel barrier for efficient spin injection to combat the notorious conductivity mismatch issue[11]. Recently, M. Farmanbar et al. predicted that insertion of BN between Co and $MoS_2$ could shift the Fermi level near the conduction band of $MoS_2$ and allow barrier free charge injection[12]. This is because h-BN can reduce the work function of Co significantly[12,13]. Most metal contacts result in p-type conduction in BP devices[14–16]. However, since h-BN is already used as an encapsulation layer for BP, depositing Co electrodes on h-BN-encapsulated BP is a potentially very simple strategy to gain access to n-type conduction, without having to utilize low work function metal contacts[17,18] or various doping methods[19]. Moreover, inert BN layer will isolate BP from contacts and hence metal-induced gap states due to strong hybridization will be eliminated[12]. Success in this approach would make BP more attractive for basic device concepts such as spin diodes[20], since h-BN/Co is ideal for spin injection and it could also allow simultaneously creating p-n junctions in the channel.

In this letter, we show experimentally that indeed, Co/h-BN contacts result in n-type conduction. Device fabrication starts with the dry transfer of ultra-thin BP crystals on ~ 20 nm h-BN crystals at 40 ºC by utilizing the PDMS method[21]. Next, encapsulating ~ 1 nm (~ 2-3 layers) thick BN layer (Fig. 1c) is deposited on BP/BN stack. All the heterostructure device preparation has been completed under inert gas environment and the BP crystal has never exposed to air. To improve the bonding between layers, the final stack is annealed at 250 ºC under high vacuum conditions for 6 hours. Our device fabrication does not involve any etching process and standard electron beam lithography technique is employed to form the contact masks. Fabrication is completed by depositing Co/Ti (30nm / 5nm) under ultra-high vacuum conditions (~ $5\times10^{-9}$ Torr) with electron beam evaporator technique. Optical and scanning electron microscopy images of device after the metallization process are shown in Fig. 1b. In this work, we studied more than 10 devices with BP thicknesses varying between 5 nm and 10 nm. We present results obtained in a ~ 5 nm (device A) and ~ 6.5 nm (device B) BP devices but all devices exhibit qualitatively very similar behavior. For comparison purposes, we also formed additional Co directly on BP without any encapsulating BN layer. This is illustrated in the schematics shown in Fig. 1a. The bias current ($I_{SD}$) of BP devices is characterized as a function of back gate voltage ($V_{BG}$), source-drain bias voltage ($V_{SD}$) and temperature under multi-terminal configuration.

Figure 1d shows the $V_{BG}$ dependence of $I_{SD}$ at room temperature for the encapsulated region in device A. It exhibits a strong n-type behavior. This is in sharp contrast to the device characteristics observed in Au contacted encapsulated device (Figure S2) and the adjacent junction with the direct Co contact where typical p-type behavior is observed (Figure 1e). The origin of this transition from p-type to n-type is due to the different interface dipoles at the contacts, and the resulting changes in the Fermi level position, as will be discussed in detail later. Simple band diagrams associated with the corresponding device characteristics for the encapsulated and non-encapsulated devices are shown in the insets of Figure 1d-e. Figure 1f shows the $V_{SD}$ dependence of $I_{SD}$ for encapsulated and non-encapsulated devices. While direct Co contacts show linear I-V relation, Co/h-BN contacts demonstrate parabolic $I_{SD}$ dependence on $V_{SD}$ confirming the tunneling nature of h-BN [22].

To understand the origin of contrasting carrier types for devices with Co/BP interfaces versus those with Co/h-BN/BP interfaces, we perform first principles density functional theory (DFT) calculations. We use the Perdew-Burke-Ernzerhof generalized gradient approximation (GGA-PBE)[23] to the exchange-correlation functional, with Grimme's dispersion corrections (PBE-D2)[24] (further details in Methods Section). Various structures including Co/xBP (x=1, 2 and 3) Co/xBN/1BP (x=1, 2 and 3) and Co/1BN/2BP interfaces were constructed, with x indicating the number of layers.

Figure 2a-b shows the atomic structure and electronic band structure of Co/xBP (x=1, 3). Interestingly, the electronic states of the first layer of BP in direct contact with Co strongly hybridizes with the Co states. The atomic structure of this layer of BP is distorted, and the projected band structure (dark blue dots in band structure; Figure 2a) differs significantly from that of pristine phosphorene (Figure S4). When more layers of BP are added, the second and third layers of BP away from Co retain the characteristic band structure of pristine phosphorene. Thus, Co with one layer of BP (Co/1BP) serves as an effective electrode. P-type contact behavior is found for Co/3BP, with $E_F$ closer to the valence band maximum ($VB_M$) of BP. When h-BN layers are added between Co and BP, the projected band structure on BP is similar to that of pristine phosphorene (Figure 2c-d, S5). This is consistent with the weak van der Waals interaction between BN and BP. Here, Co/1BN serves as the effective electrode, with the remaining BN layers serving as a tunneling barrier. In contrast to the Co/BP interfaces, our calculations show that Co/xBN/BP interfaces are n-type contacts, in agreement with our experimental findings.

The contrasting band alignment at Co/BP and Co/xBN/BP interfaces arises from the large differences in work function between the effective electrodes, Co/1BP and Co/1BN, respectively (Figure 3c). The work function of Co/1BP is 4.47 eV, very close to that computed for bare Co (4.86 eV). In contrast, the work function of Co/1BN is much smaller (3.63 eV), and additional BN layers do not change this value significantly. (We note that while PBE-D2 was chosen because it gives better structural parameters for BP, the computed work functions are systematically underestimated compared to LDA[25]. Changes in work function $\Delta W$, when BP and BN layers are added to Co, are directly proportional to the interface dipole moment $d$ per unit area, $\Delta W = ed/\varepsilon_0$[13]. From Figure 3a, we can see that there is a significant depletion of electrons adjacent to BN and a large accumulation of electrons close to Co at the Co/1BN interface. This large interface dipole arises from the adsorbed BN layer pushing electrons towards the Co substrate because of Pauli repulsion (called the pillow effect), and decreases the work function by 2.04 eV ($d = 0.0112$ e/Å). At the same time, the slight buckling of the adsorbed BN sheet, with the B and N atoms moving towards and away from the Co surface respectively, gives rise to another dipole moment that cancels the interface dipole moment above. This gives a net decrease of the Co work-function by 1.18 eV upon BN adsorption, consistent with Figure 3c.

Why then is the work function of Co/1BP so similar to Co? Firstly, BP is non-polar, and does not contribute any dipoles to the surface. Secondly, charge transfer from Co to BP results in an interface dipole that cancels the interface dipole from the pillow effect (Figure 3b). In contrast to Co/1BN (Figure 3a), the electron charge redistribution plot in Figure 3b shows electron depletion close to both BP and Co at the BP/Co interface, and a very large electron accumulation between these two regions. Electron depletion close to BP arises from the pillow effect. Electron depletion close to Co can be attributed to electron transfer from the occupied minority spin Co states to BP conduction bands, owing to the overlap between these states as shown in the band structure of Figure 2a. The resulting dipole moment of the Co/1BP interface is very small ($d = 0.0018$ e/Å, pointing from Co to BP), giving a work function decrease of 0.32 eV, in good agreement with Figure 3c. Crucially, the band gap in BN is so large that no such charge transfer occurs, allowing for a large reduction in the work function. Interestingly, in the case of graphene on Co/Ni surfaces, charge transfer also takes place, but in the opposite direction, thus giving rise also to a large decrease in work function[26]. We note that it has been previously predicted that h-BN does not change the work function of Au significantly, consistent with our findings in Figure S2[13].

Now we discuss the detailed transport characterization of Co/h-BN contacted device. Figure 4-a shows the room temperature $V_{BG}$ dependence of $I_{SD}$ at fixed $V_{SD}$ values for the encapsulated device B. At a fixed $V_{BG}$, *on* current increases exponentially as $V_{SD}$ increases due to tunneling nature of the contacts (See also Figure 4b). We carried out a detailed study of $I_{SD}/V_{SD}$ as a function of temperature in order to confirm the tunneling nature of our contacts. We study the temperature dependency of $I_{SD}$-$V_{SD}$ data by using the thermionic emission model. This model analyzes the charge injection into the BP through a Schottky barrier by using[27–29]

$$I_{SD} = AA^*T^{1.5}exp\left[-\frac{e}{k_BT}\left(\Phi_B - \frac{V_{SD}}{n}\right)\right]$$

where $A$ is the contact area, $A^*$ is the 2D Richardson constant, $e$ is the electron charge, $k_B$ is the Boltzmann constant and $\Phi_B$ is the Schottky barrier height and $n$ is the ideality factor. As shown in Figure 4b, $I_{SD}$ is insensitive to temperature therefore the extracted values are small and indeed negative. This confirms the absence of any Schottky barrier height[30]. Note that direct Co contacts to BP results in Schottky barrier height as high as 206 meV[28]. Temperature dependence of conductivity (σ) vs $V_{BG}$ at a fixed $V_{SD}$ value shows the insulator-to-metal transition (MIT) under 2T geometry (Figure 4c), where metallic behavior is observed starting at $V_{BG}$ ~ 13V, with a critical conductivity of ~ $e^2$/h. This is surprising as MIT in 2D materials under 2T geometry could only be achieved in ion liquid gated devices at very high carrier concentrations[31]. Our low resistance, barrier free contacts enables us to observe MIT at such low bias voltages with conventional 300 nm SiO$_2$ + ~ 20 nm BN solid gates. Next, we obtain the contact resistance by following the protocol shown in Ref.[32]. Measurements are performed using the four and two terminal geometries (4T and 2T). The extracted R$_C$ depends on both V$_{BG}$ and V$_{SD}$. As shown in Figure 4d, R$_C$ (1/2 of the extracted total contact resistance by assuming that contacts are symmetric) decreases from ~ 6.7 kΩ to ~ 4.5 kΩ while V$_{BG}$ is tuned from 10 V to 62.5 V, respectively. Since I$_{SD}$/V$_{SD}$ of our contacts show parabolic dependence due to tunneling nature, we observe that R$_C$ depends on V$_{SD}$ as well and it decreases from ~ 6.6 kΩ to ~ 4.5 kΩ as V$_{SD}$ increases from 50 mV to 0.5 V, respectively at V$_{BG}$=62.5V. The extracted contact resistance is superior to those Ohmic contacts achieved in encapsulated devices using graphene contacts where the lowest achieved R$_C$ is 11.2 kΩ[3].

With the Schottky barrier-free contacts, it is possible to access the intrinsic properties of BP devices. Here, we calculate the field effect mobilities measured under 4T and 2T geometries by using $\mu = \frac{L}{W}\frac{1}{C_S}\frac{1}{V_{SD}}\frac{dI_{SD}}{dV_{BG}}$ where $\mu$ is the field effect mobility, *L* and *W* are the length and width of the measured junction, respectively, and $C_S$ is the substrate capacitance. Figure 4e shows the temperature dependence of mobility extracted near n ~ 3x10$^{12}$ cm$^{-2}$. Mobility measured under 4T geometry excludes any contact contributions and it is nearly temperature independent at low temperatures[33]. It decreases monotonically from ~ 4,190 cm$^2$/Vs to ~ 245 cm$^2$/Vs with increasing the temperature from 40 K to 300 K. Such temperature dependence of mobility is similar to what has been extracted for hole conduction channel at the previous studies[7]. We conclude that the charge transport is limited due to impurities at low temperatures as the mobility is temperature independent. Here, we take advantage of using h-BN as a substrate. Utilizing h-BN as a substrate reduces the impurity dependent scattering compared to the conventional SiO$_2$ substrates[8] and hence higher mobilities at low temperatures are achieved. At higher temperatures phonons start to dominate the momentum relaxation. Next, we discuss the mobility of BP measured under 2T geometry (Bottom panel of Figure 4e). We observe that mobility increases as temperature decreases and then becomes temperature independent - similar characteristics to those observed under 4T measurement configuration. This is indeed in sharp contrast to previously characterized graphene[3] and Ti/Au[16] contacted BP devices where mobility decrease up to 80% as temperature is reduced. The improvement in our study is due to the tunneling characteristic of contacts and the Schottky barrier free charge injection. Here, we note that the magnitude of mobility measured under 2T geometry is an order of magnitude smaller than 4T case. This is due to the presence of finite contact resistance which indeed constitute the ~ %74 of total 2T resistance at high gate voltages (Figure 4d)[33].

Previously, n-type contact to BP has been achieved by carefully engineering the contact material[17]. However this effect depends on the crystal thickness. Our mechanism completely relies on Co-BN interface and it results in n-type doping, independent of the crystal thickness. Inserting BN layer between contact and BP also prevents the formation of mid-gap states in sharp contrast to the direct contacts case. This allows Schottky barrier free charge injection into BP[12]. Besides serving as an encapsulating layer to prevent the BP from degradation[3], our contact scheme is also crucial for realizing spin dependent transport phenomena in BP[34].

Finally, we fabricate a p-n junction by utilizing the reduced work-function of Co upon insertion of a thin layer of h-BN underneath. Here, Co/h-BN and Co contacts are employed as source and drain electrodes, respectively (Figure 5a). Figure 5b shows the 2D color fan plot of $I_{SD}$ as a function of $V_{BG}$ and $V_{SD}$ and line cuts at both axes are also plotted in Figure 5c. $I_{SD}$-$V_{BG}$ strongly depends on $V_{SD}$. We observe completely p-type behavior at $V_{SD}$=4V. Transport characteristics of device change from p-type to ambipolar and eventually to n-type dominant conduction as $V_{SD}$ is tuned from 4 V to -4V. Rectification ratio and current polarity depends on the $V_{BG}$ and highest rectification ratio achieved is 20. We note that higher ratios have been obtained for BP previously by combining it with n-type $MoS_2$[35]. However, our device architecture demonstrates an alternative way of constructing lateral p-n junctions by using only BP as an active material. Its performance can be further improved by optimizing the separation between source and drain electrodes. In future experiments, this approach could be integrated into vertical devices to form sharp vertical p-n junctions if the bottom part of the crystal is contacted to a p-type contact.

In conclusion, we investigated the transport characteristics of ultra-thin BP crystals contacted with van der Waals bonded Co/h-BN. Our combined experimental and theoretical study reveals that Co/h-BN contacted BP transistors show n-type characteristics due to the large interface dipole at Co/h-BN, which decreases the work function significantly compared to Co/1BP. Low contact resistances and Schottky barrier free charge injection into BP are obtained using Co/h-BN contacts. These contacts allow us to observe insulator-to-metal transitions under 2T geometry with conventional $SiO_2$ substrates, and to probe intrinsic electron mobilities of BP (~4,200 $cm^2$/Vs).

Encapsulating h-BN layer has enabled the exploration of many interesting air sensitive 2D crystals last two years and it will remain to be a key component for the discovery of new 2D crystals. We believe that this van der Waals contacted Co/h-BN contacts is potentially very simple strategy to have an access to intrinsic properties of these crystals while keeping them air-stable. From a technology point of view, such contacts could make 2D materials also attractive for basic device concepts such as spin diodes, since they can be used as spin injector and simultaneously control n and p-type conduction in BP.

**Note**: During the preparation of this manuscript, we became aware of a related work on the electronic transport in $MoS_2$-based heterostructures where h-BN/Co contacts are similarly utilized[36].

Theoretical Methods

The first principles calculations are performed using a plane wave basis set at the level of density functional theory (DFT). The Perdew-Burke-Ernzerhof generalized gradient approximation (GGA-PBE)[23] to the exchange-correlation functional is employed in projector-augmented wave (PAW)[37] potentials as implemented in the Vienna ab initio simulation package (VASP)[38]. The van der Waals interactions are taken into account through Grimme's dispersion corrections (PBE-D2) for all layered structures[24]. We have also computed the HSE06 band structure for pristine monolayer and bilayer BP. Compared to the PBE band structure, the HSE06 calculations shift the valence band minimum down by 0.42 eV and 0.36eV in monolayer and bilayer BP, respectively, and shifts the conduction band maximum up by 0.26 and 0.25eV in monolayer and bilayer BP, respectively. These shifts will not change the qualitative predictions of p-type and n-type contacts in Co/BP and Co/xBN/BP interfaces, respectively. The (111) surface of Co has a similar lattice constant (~2.49Å) as the hexagonal boron nitride (h-BN) layer, with a lattice mismatch of less than 0.1%. For the Co/h-BN interface with black phosphorus (BP), a 1x3 supercell of BP shares the same lattice as a Co/h-BN cell that is rotated by 30° with lattice vectors $\mathbf{b_1} = \mathbf{a_1}+\mathbf{a_2}$, $\mathbf{b_2} = 4\mathbf{a_1}-4\mathbf{a_2}$, where $\mathbf{a_1}$, $\mathbf{a_2}$ are the original lattice vectors of the primitive Co/h-BN surface. This leads to a ~1.5% compressive strain in Co/h-BN. The Co(111) surface is simulated by five atomic layers of Co. A vacuum layer at least 16 Å thick is added between the slabs to prevent interactions. A kinetic energy cut-off of 450 eV is used for the plane wave basis set, and the Brillouin Zone is sampled by a Monkhorst-Pack k-point mesh of 11×5×1.

All atoms are relaxed with a force convergence criterion of 0.01 eV Å$^{-1}$. In the work function calculations, the unit cells of Co, Co/1BP, Co/xBN (x=1, 2, and 3) are used and a much denser BZ k-point sampling of 36×36×1 is used to get converged results, with the convergence threshold for energy set to $10^{-6}$ eV in the self-consistent calculation. To analyze the interface surface dipoles that impact the work function of these effective electrodes, a charge density difference $\Delta n$ analysis is carried out, where $\Delta n$ is defined by the difference between the electronic charge density of the effective electrode (Co/1BP or Co/1BN), and the electronic charge density of individual components (Co and 1BP, or Co and 1BN, respectively). $\Delta n$ is further averaged in the plane of the surfaces, so that $\Delta n$ is plotted as a function of z, the direction normal to the surfaces. For example, at the interface of the Co/1BN electrode, $\Delta n(z) = n_{Co/1BN}(z) - n_{Co}(z) - n_{1BN}(z)$. The change in work function caused by the interface dipole is given by $\Delta W = ed/\varepsilon_0$, where $d$ is the magnitude of the dipole moment per unit area and $\varepsilon_0$ is the vacuum permittivity[13].

## AUTHOR INFORMATION

Author Contributions

## ACKNOWLEDGEMENTS

B. Ö. would like to acknowledge support by the National Research Foundation, Prime Minister's Office, Singapore, under its Medium Sized Centre Programme and CRP award "Novel 2D materials with tailored properties: beyond graphene" (Grant number R-144-000-295-281) and Competitive Research Programme (CRP Award No. NRF-CRP9-2011-3). L.X. and S.Y.Q. acknowledge support from the Singapore National Research Foundation under grant NRF-NRFF2013-07. Computations were performed on the NUS Graphene Research Centre cluster. We acknowledge support from the Singapore National Research Foundation, Prime Minister's Office, under its medium-sized centre program. K.W. and T.T. acknowledge support from the Elemental Strategy Initiative conducted by the MEXT, Japan and JSPS KAKENHI Grant Numbers JP15K21722 and JP25106006.

SUPPORTING INFORMATION

The Supporting Information is available free of charge on the ACS Publications website.

BP height characterization; device characteristics of BP contacted to h-BN/Au; device characteristics of BP contacted to h-BN/Co; field effect mobility of BP contacted to h-BN/Co; band structure of isolated monolayer BP; atomic structure and electronic band structure of Co/1h-BN/1BP; majority band structures.

Figure Captions:

**Figure 1:** (a) Device schematics with measurement configuration. (b) Optical and scanning electron microscopy images of heterostructure device which is made of a bottom ~ 20 nm BN, ~ 6.5 nm BP and ~ 1 nm (3 layers) top BN layers. (c) Atomic force microscopy image of encapsulating thin BN layer. Color scale bar is 2 nm. (d-e) $V_{BG}$ dependence of $I_{SD}$ at fixed $V_{SD}$ = 1V in the encapsulated and non-encapsulated devices, respectively. Note that the encapsulated device has hysteresis free transport characteristic even at RT unlike the non-encapsulated device. (f) $V_{SD}$ dependence of $I_{SD}$ for Co/BN and Co contacted junctions.

**Figure 2:** Atomic structure and electronic band structure of (a) Co/1BP (b) Co/3BP, (c) Co/3BN/1BP, (d) Co/1BN/2BP. The Fermi level is set at zero eV. The colored dots in the band structure denote the projections onto different parts of the system, as indicated in the schematics. The sizes of the dots are proportional to the weights of the projections of the electronic states onto the specific atoms. For example, the dark blue dots in Figure 2a represent the band structure projected onto BP. The minority spin band structure is shown. The majority band structure is provided in the Supplementary Information (Figure S6) and does not change the conclusions.

**Figure 3**: Plane averaged electronic charge density difference $\Delta n(z)$ for (a) Co/1BN and (b) Co/1BP. Dashed lines in (a) and (b) indicate average z-positions of atomic planes (note that BP has two atomic planes while BN has one). Interface dipoles caused by the pillow effect and charge transfer are indicated schematically. The interface dipole from buckling of the BN sheet is omitted in this schematics. (c) Calculated work-functions for Co, Co/1BP and Co/ xBN.

**Figure 4**: (a) $V_{BG}$ dependence of $I_{SD}$ at different $V_{SD}$ values measured under two terminal configuration. (b) Temperature dependence of $I_{SD}/V_{SD}$ at $V_{BG} = 25$ V. (c) Temperature dependence of $\sigma/V_{BG}$. (d) $V_{BG}$ dependence of 2T, 4T device and extracted total contact resistances. Inset shows the $V_{SD}$ dependence of $R_C$. (e) Four-and two-terminal field effect mobilities of h-BN/Co contacted BP device as a function of temperature. For comparison purposes, performance of graphene contacted BP (similar thickness) measured in Ref.3.

**Figure 5**: (a) Schematics for the device structure. (b) 2D color fan plot of $I_{SD}$ as a function of $V_{SD}$ and $V_{BG}$. (c) Replotted version of (b) for better clarity. Inset shows the $V_{BG}$ dependence of $I_{SD}$ at fixed $V_{SD}$ values of 4V (black line) and -4V (red line).

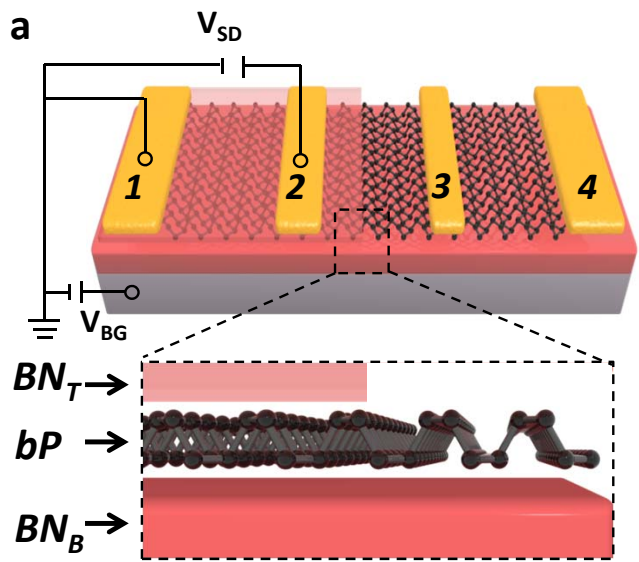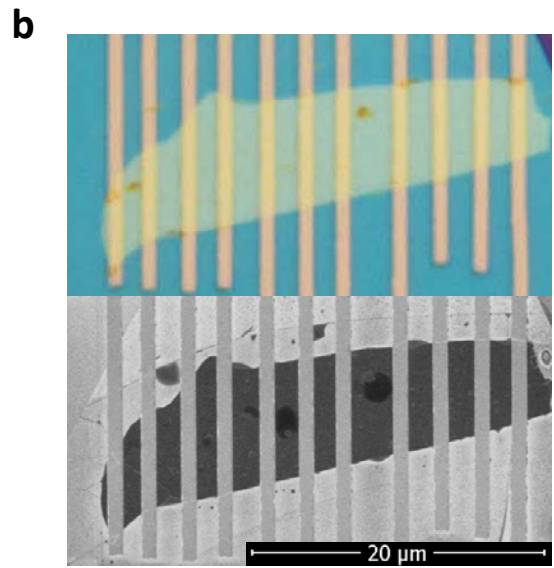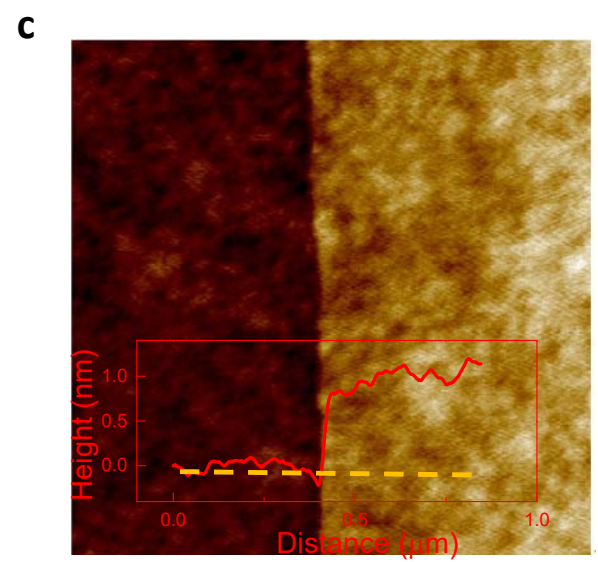
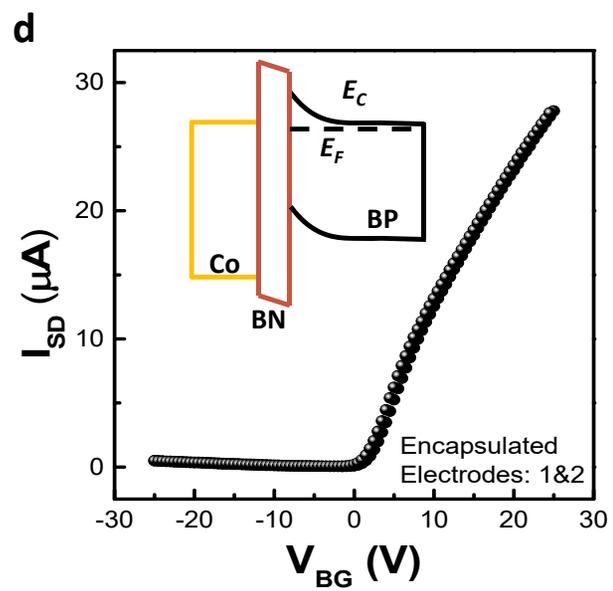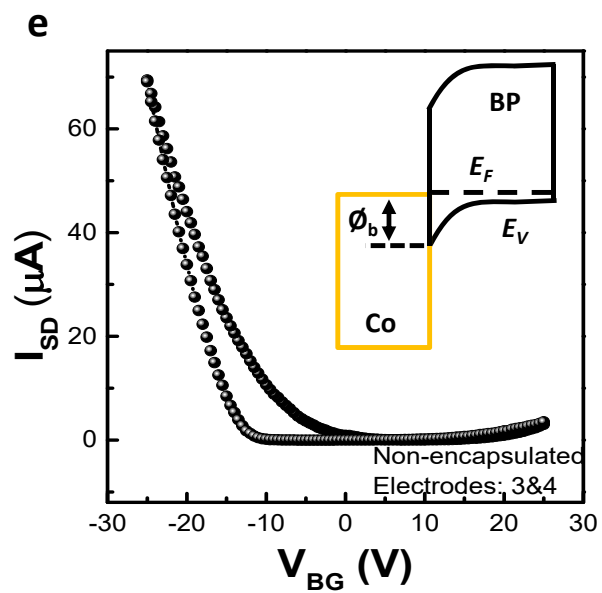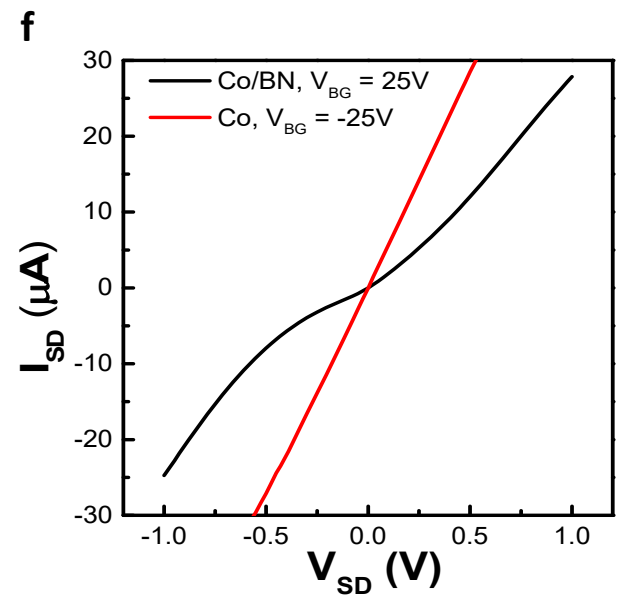

**Figure 1**

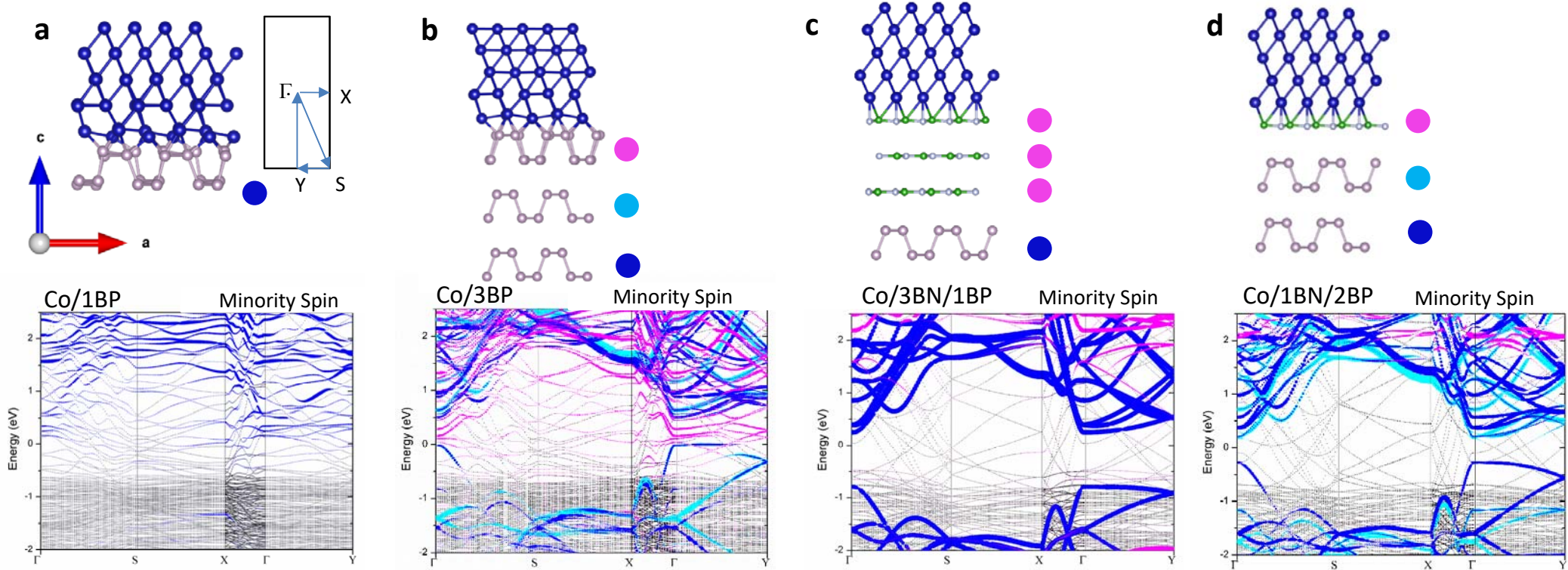

**Figure 2**

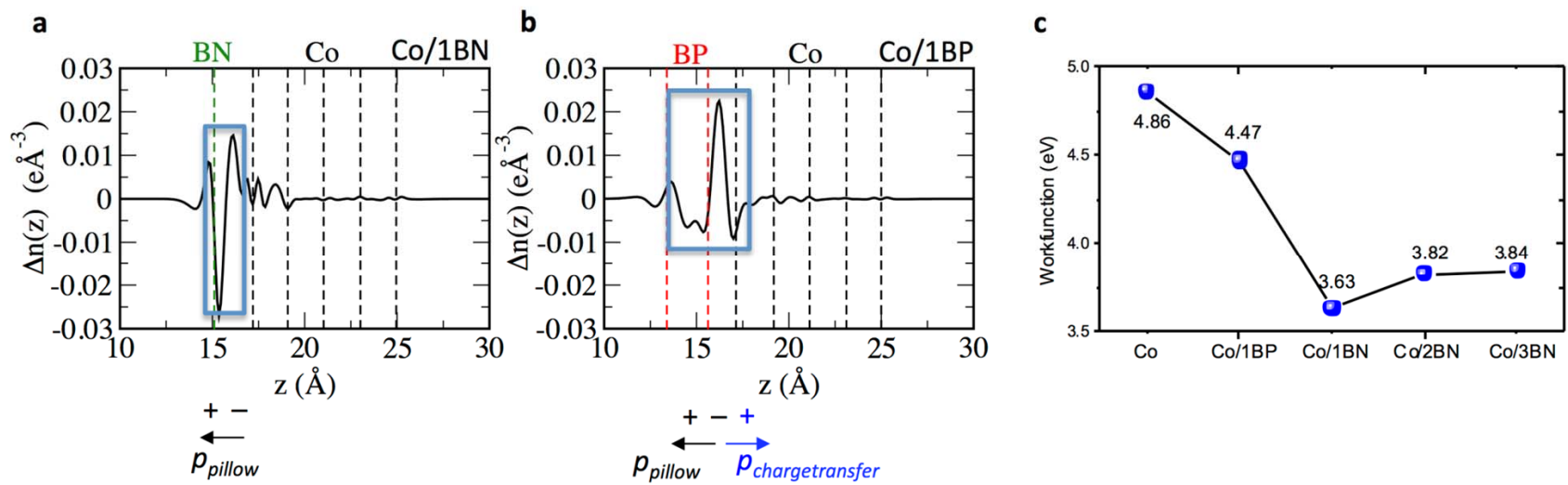

**Figure 3**

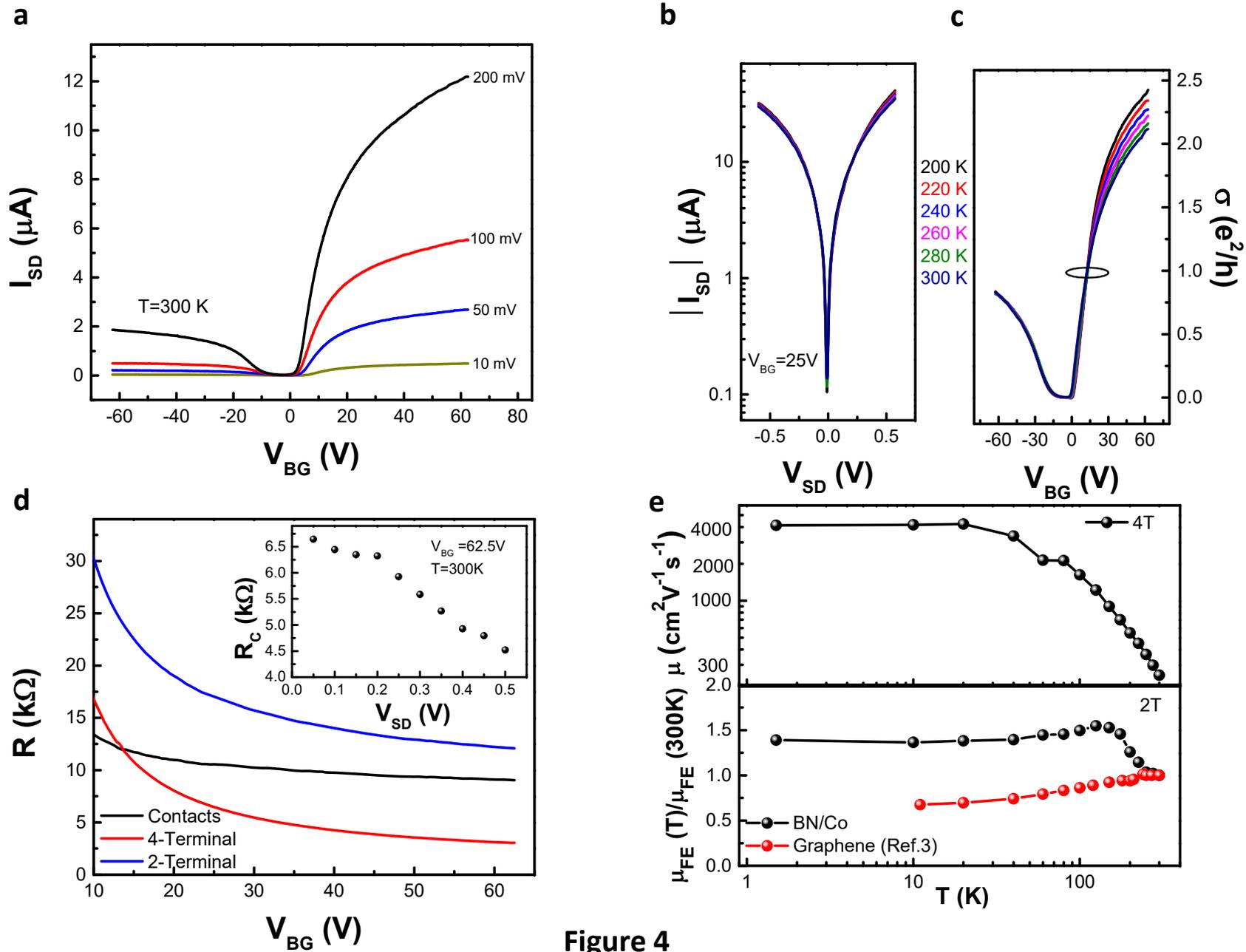

**Figure 4**

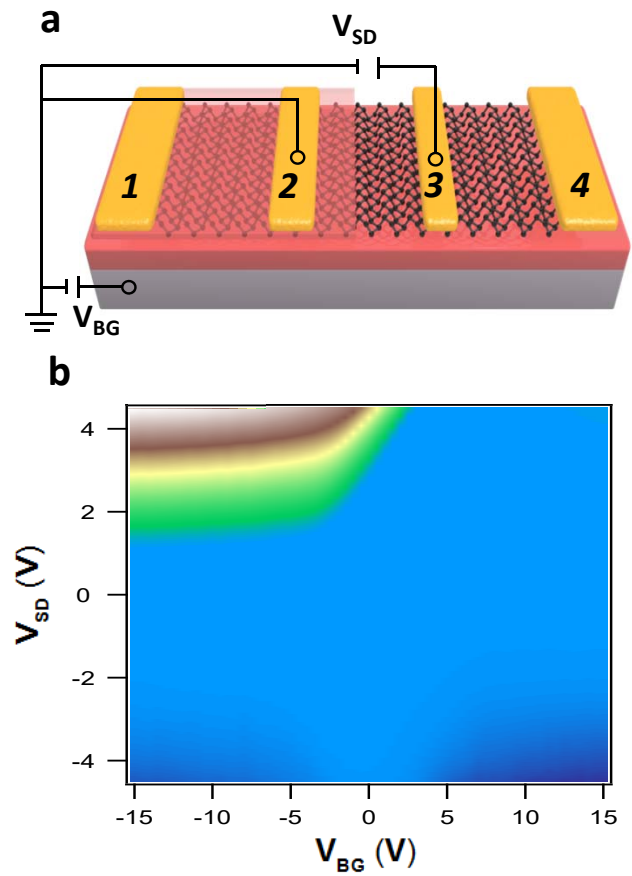
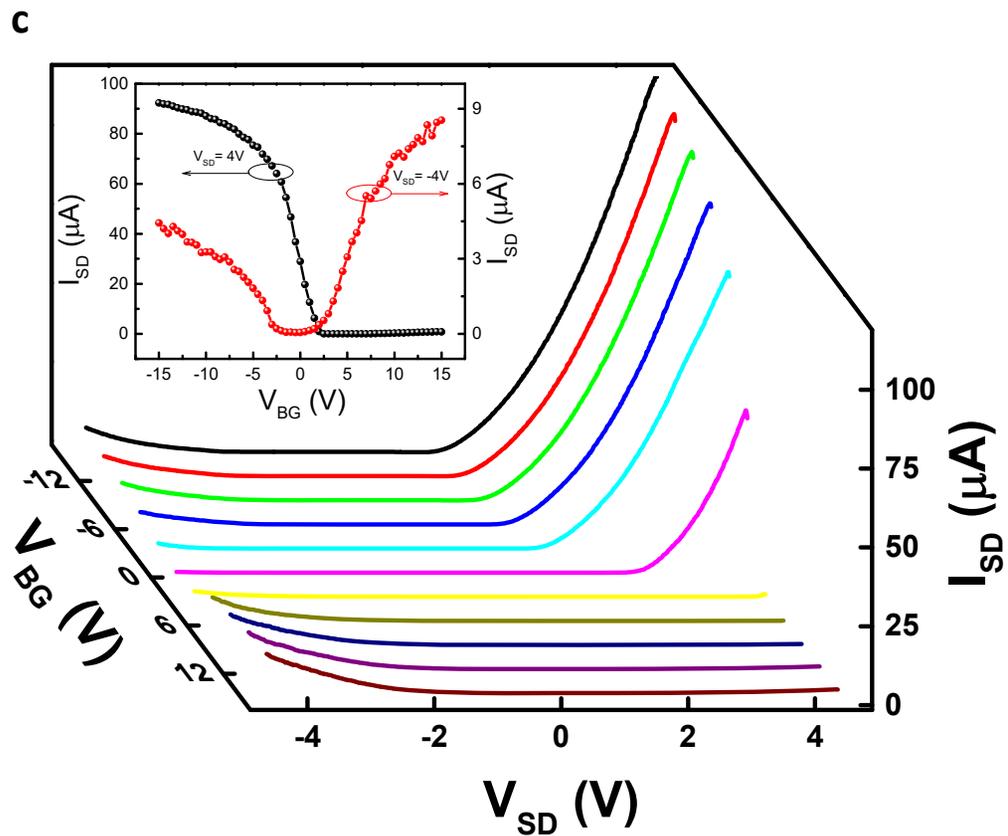

**Figure 5**